 \documentclass[]{jfm}
\usepackage{graphicx}
\usepackage{amsmath}%
\usepackage{subfig}
\usepackage{psfrag}
\usepackage[usenames, dvipsnames]{color}
\usepackage{float}
\usepackage{natbib}
\usepackage{pifont}
\usepackage{siunitx}
\ifCUPmtlplainloaded \else
  \checkfont{eurm10}
  \iffontfound
    \IfFileExists{upmath.sty}
      {\typeout{^^JFound AMS Euler Roman fonts on the system,
                   using the 'upmath' package.^^J}%
       \usepackage{upmath}}
      {\typeout{^^JFound AMS Euler Roman fonts on the system, but you
                   dont seem to have the}%
       \typeout{'upmath' package installed. JFM.cls can take advantage
                 of these fonts,^^Jif you use 'upmath' package.^^J}%
      }
  \else
  \fi
\fi

\ifCUPmtlplainloaded \else
  \checkfont{msam10}
  \iffontfound
    \IfFileExists{amssymb.sty}
      {\typeout{^^JFound AMS Symbol fonts on the system, using the
                'amssymb' package.^^J}%
       \usepackage{amssymb}%
       \let\le=\leqslant  
         
      }{}
  \fi
\fi

\ifCUPmtlplainloaded \else
  \IfFileExists{amsbsy.sty}
    {\typeout{^^JFound the 'amsbsy' package on the system, using it.^^J}%
     \usepackage{amsbsy}}
    {\providecommand\boldsymbol[1]{\mbox{\boldmath $##1$}}}
\fi

\newcommand{\diff}{\mathrm{d}}

\begin{document}

\title{On the role of secondary motions in turbulent square duct flow}

\author[D. Modesti, S. Pirozzoli, P. Orlandi, F. Grasso]{DAVIDE MODESTI$^2$, SERGIO PIROZZOLI$^1$ \\ PAOLO ORLANDI$^1$ AND FRANCESCO GRASSO$^{2}$}

\affiliation{$^1$Dipartimento di Ingegneria Meccanica e Aerospaziale,
Sapienza Universit\`a di Roma \\ Via Eudossiana 18, 00184 Roma,
Italy \\ $^2$ Cnam-Laboratoire DynFluid, 151 Boulevard de L'Hopital, 75013 Paris\\[\affilskip]}

\maketitle

\begin{abstract}
We use a direct numerical simulations (DNS) database for turbulent flow in a square duct up to bulk Reynolds
number $\Rey_b=40000$, to quantitatively analyze the role of secondary motions on the mean flow structure.
For that purpose we derive a generalized form
of the identity of Fukagata, Iwamoto and Kasagi (FIK), which allows to quantify the effect of cross-stream
convection on the mean streamwise velocity, wall shear stress and bulk friction coefficient. Secondary motions
are found to contribute for about $6\%$ of total friction, and to act as a self-regulating mechanism of turbulence
whereby wall shear stress nonuniformities induced by corners are equalized, and 
universality of the wall-normal velocity profiles is established. 
We also carry out numerical experiments 
whereby the secondary motions are artificially suppressed, in which case their equalizing role 
is partially taken by the turbulent stresses.
\end{abstract}

\section{Introduction}

Flows within ducts with square cross section are the
simplest prototype of internal flows with two-dimensional mean flow statistics,
and they are common in many engineering applications,
such as heat exchangers, turbomachinery, nuclear reactors and
water draining and ventilation systems.
These flows are characterized by the appearance of secondary motions, first advocated
by~\citet{prandtl_27} and \citet{nikuradse_30}
to explain bulging of the streamwise velocity isolines towards the duct corners.
Secondary motions in square duct flow consist of an array of eight counter-rotating vortices,
bringing high-momentum fluid from the core towards the corners.
Early quantitative measurements~\citep{brundrett_64,gessner_65} showed that the 
intensity of the secondary eddies is about 1-2\% of the bulk velocity, 
as also confirmed in much later numerical studies~\citep{gavrilakis_92,pinelli_10,vinuesa_14}.
Although their origin is not fully understood, theoretical analysis made it clear that they
are associated with gradients of the secondary turbulent stresses, absent in canonical 
pipe flow~\citep{speziale_82}. A simple model for the structure of the secondary eddies has been recently 
proposed by the present authors~\citep{pirozzoli_18}, supported by DNS data at computationally high
Reynolds number. 

In most studies it is stated that, despite their weakness, secondary motions significantly affect 
the structure of the mean velocity field, and hence have important influence on the
distribution of wall friction and on particle dispersion.
However, to the best of our knowledge, this effect has never been quantified, mainly because of 
difficulties in isolating the effect of mean cross-stream convection from turbulence.
This is the goal of the present paper, in which we attempt to
quantify the role of secondary motions in the redistribution of momentum based on DNS data.
For that purpose we derive a generalized form of the FIK identity~\citep{fukagata_02}
which allows to separate the effect of mean convection on both local and mean friction.
We also carry out numerical experiments whereby the
mean cross-stream flow is artificially suppressed to
further clarify its effect on the mean flow statistics.

\section{Methodology}\label{sec:methodology}

\begin{table}
\centering
\begin{tabular}{lccccccccccc}
\hline
\hline
 Case & $\Rey_b$ & $\Rey_{\tau}^*$ & $C_f\times10^{3}$ & $\Delta C_f(\%)$ & $N_x$ & $N_y$ & $N_z$ & $\Delta x^*$ & $\Delta z^*$ & $\Delta y_w^*$ &${\Delta t}_{av}^* u^*_\tau/h$ \\
\hline
 A &$4410$ & $150$  & $9.26$ & 0 & $512$  & $128$ & $128$ & $5.6$ & $3.0$  & $0.55$ & $2120$  \\
 B &$7000$&  $227$  & $8.41$ & 0 & $640$  & $144$ & $144$  & $6.6$ & $4.8$ & $0.51$ & $1607$  \\
 C &$17800$& $519$  & $6.80$ & 0 & $1024$ & $256$ & $256$ & $9.5$ & $6.3$  & $0.53$ & $1387$  \\
 D &$40000$& $1055$ & $5.57$ & 0 &  $2048$ & $512$ & $512$ & $9.6$ & $6.4$  & $0.60$ & $531$   \\
\hline
 A0 &$4410$ & $154$  & $9.76$ & 5.4  &  $512$  & $128$ & $128$ & $5.7$ & $3.4$  & $0.56$ & $1049$  \\
 B0 &$7000$&  $230$  & $8.64$ & 2.7  &  $640$  & $144$ & $144$  & $6.7$ & $4.5$ & $0.52$ & $120$  \\
 C0 &$17800$& $511$  & $6.59$ & -3.1 &  $1024$ & $256$ & $256$ & $9.4$ & $6.1$  & $0.65$ & $144$  \\
 D0 &$40000$& $1038$ & $5.39$ & -3.2 &  $2048$ & $512$ & $512$ & $9.55$ & $6.37$  & $0.59$ & $29$   \\
\hline
\end{tabular}
\caption{Flow parameters for square duct DNS.
The box dimension is $6\pi h \times 2h \times 2h$ for all flow cases.
$\Rey_b = 2 h u_b / \nu$ is the bulk Reynolds number, and
$\Rey_{\tau}^* = h u_{\tau}^* / \nu$ is the friction Reynolds number.
$\Delta x$ is the mesh spacing in the streamwise direction, and
$\Delta z$, $\Delta y_w$ are the maximum and minimum mesh spacings
in the cross-stream direction, all given in global wall-units, $\delta_v^*=\nu/u_{\tau}^*$.
${\Delta t}_{av}^*$ is the effective averaging time interval.
Flow cases denoted with the 0 suffix are carried out by suppressing the secondary motions.
$\Delta C_f(\%)$  denotes   the percent  friction  increase/reduction
of flow cases with suppression of secondary motions with respect to the reference ones.
}

\label{tab:testcases_duct}
\end{table}
This study mainly relies on the DNS database developed by \citet{pirozzoli_18}.
The compressible Navier-Stokes equations are solved using a fourth-order co-located
finite difference solver~\citep{modesti_16impl}, and integrated over very long time intervals 
to guarantee statistical convergence.
Four simulations have been carried out at low bulk Mach number, $M_b=0.2$, in the range of bulk
Reynolds number $\Rey_b=4410-40000$ ($\Rey_b=2hu_b/\nu$, where $h$ is the duct half side length, $u_b$ the bulk velocity and $\nu$ the fluid kinematic viscosity), and labeled as A-D in table~\ref{tab:testcases_duct}. 
A thorough validation was carried out by~\citet{pirozzoli_18},
showing good agreement between the new dataset and 
previous DNS studies~\citep{pinelli_10,vinuesa_14}.
For the sake of clarity, the velocity components in the streamwise
and wall-normal directions are denoted as $u$, $v$ and $w$, respectively, and the
overline symbol is used to indicate statistical averages in the streamwise direction and in time.
In the following both local wall units (based on the local friction velocity, $u_\tau=\sqrt{\tau_w/\rho}$, with $\tau_w$ the local wall shear stress)  and global wall units (based on the mean shear stress) are used and are denoted with ${(.)}^+$ and ${(.)}^*$, respectively.
 
Additional numerical experiments have been carried out here
at the same bulk Reynolds numbers as the baseline cases (and denoted with the 0 suffix in table~\ref{tab:testcases_duct}), 
by artificially suppressing the
secondary motions. For that purpose
we force the streamwise-averaged cross-stream velocity components to have zero mean by setting
\begin{subequations}
 \begin{align}
  v(x,y,z,t) &\rightarrow v(x,y,z,t) - \overline{v}^x(y,z,t)\\ 
  w(x,y,z,t) &\rightarrow w(x,y,z,t) - \overline{w}^x(y,z,t), 
 \end{align}
\end{subequations}
at each Runge-Kutta sub-step,
where $\overline{(.)}^x$ denotes the streamwise averaging operator.
Note that this procedure is independent from the streamwise length of the
computational domain or the Reynolds number, as by definition the modified velocities have zero mean, 
hence the secondary motions are suppressed to machine accuracy.
%We applied the procedure both at each time step and at each Runge-Kutta sub-step,
%without apparent differences in the results, but the latter is preferred as it guarantees
%that intermediate velocities are also free from secondary flows, which seems more consistent.}
%
Although the modified velocity field does not strictly satisfy the
Navier-Stokes equations, the resulting statistics may be instructive to
understand the role of the secondary motions. 
Previous studies~\citep{pinelli_10,vinuesa_14} highlighted
the need for very long averaging time, thus we have collected flow statistics for equivalent times ${\Delta t}_{av}^*={\Delta t}_{av} L_x/(6h)$,
which are considerably longer than in classical plane channel flow. On the other hand, we have found that numerical 
experiments carried out without secondary motions require much shorter time averaging interval as a results of the
suppression of the slow dynamics associated with the secondary currents.
\section{The role of secondary motions} \label{sec:role}

In order to quantify the effect of secondary motions on momentum redistribution,
we derive a generalized version of the FIK identity~\citep{fukagata_02} for ducts
with arbitrary shape.
For that purpose we consider the mean streamwise momentum balance equation, namely
\begin{equation}
\nu\nabla^2\overline{u} = \nabla \cdot\boldsymbol{\tau}_C + \nabla\cdot\boldsymbol{\tau}_T
- \overline{\Pi} ,
\label{eq:mmb}
\end{equation}
where $\overline{u}$ is the mean streamwise velocity, $\boldsymbol{\tau}_C=\overline{u}\,\overline{\mathbf{u}}_{yz}$
is associated with mean cross-stream convection (hence, with the secondary motions), $\boldsymbol{\tau}_T=\overline{u'\mathbf{u'}}_{yz}$ is associated with turbulence convection,
$\mathbf{u}_{yz}=\left({v},{w}\right)$ is the cross-stream velocity vector, $\overline{\Pi}=P \tau_w^*/ (\rho A)$ is the driving pressure gradient, and $A$ and $P$ are the duct cross-sectional area and perimeter, respectively.
Equation~\eqref{eq:mmb} may be interpreted as a Poisson equation for the mean streamwise velocity, with source terms
$\nabla\cdot\boldsymbol{\tau}_T$, $\nabla\cdot\boldsymbol{\tau}_C$ and $\overline{\Pi}$ obtained from the DNS dataset.
Hence, the solution of equation \eqref{eq:mmb} may be cast as the superposition of three parts, 
namely $\overline{u}=\overline{u}_V+\overline{u}_T+\overline{u}_C$, with
\begin{equation} 
\nu\nabla^2\overline{u}_V=-\overline{\Pi}, \quad \nu\nabla^2\overline{u}_T=\nabla\cdot\boldsymbol{\tau}_T, \quad \nu\nabla^2\overline{u}_C=\nabla\cdot\boldsymbol{\tau}_C,
\label{eq:fik_vel}
\end{equation} 
with homogeneous boundary conditions,
where $\overline{u}_V$, $\overline{u}_T$, and $\overline{u}_C$ denote the viscous, turbulent, and convective contributions to the mean streamwise velocity field. 
The bulk velocity in the duct
may accordingly be evaluated as
\begin{equation}
u_b=\frac {\overline{\Pi} A}{\nu} {u_b}_1+{u_b}_T+{u_b}_C, \quad {u_b}_X = \frac{1}{A} \int_A\overline{u}_X\mathrm{d}A, \label{eq:ubulk}
\end{equation}
where we have introduced the unitary velocity field $u_1$ defined as solution of $\nabla^2 u_1 = -1/A$, which  
by construction is only a function of the duct geometry, and whereby the viscous velocity
field may be expressed as $u_V=\overline{\Pi}A u_1 / \nu$.
Inserting the friction coefficient $C_f = \overline{\Pi} D / (2 u_b^2)$ (where $D=4 A/P$ is the duct hydraulic diameter) 
into equation~\eqref{eq:ubulk} one obtains
\begin{equation}
C_f=\frac{2}{{u_b}_1 Re_P}\left(1-\frac{{u_b}_T}{u_b}-\frac{{u_b}_C}{u_b}\right)={C_f}_V+{C_f}_T+{C_f}_C,
\label{eq:fik}
\end{equation}
where $\Rey_P=u_b P / \nu$ is the bulk Reynolds number based on the duct perimeter.
Equation~\eqref{eq:fik} clearly shows that the friction coefficient may be regarded as
the sum of the contributions of viscosity, turbulence, and convection (labeled as ${C_f}_X$).
We note that this formalism has additional generality than the classical FIK identity 
and its extensions~\citep{peet_09,jelly_14},
and it further allows to isolate the effects of mean convection, 
turbulent and viscous terms to the wall shear stress distributions along the duct perimeter, resulting from
\begin{equation}
{\tau_w}_V = \rho\nu\left.\frac{\partial \overline{u}_V}{\partial n}\right\rvert_w, \quad 
{\tau_w}_T = \rho\nu\left.\frac{\partial \overline{u}_T}{\partial n}\right\rvert_w, \quad
{\tau_w}_C = \rho\nu\left.\frac{\partial \overline{u}_C}{\partial n}\right\rvert_w, \label{eq:tauw}
\end{equation}
where $n$ is the wall-normal direction. Regarding equation~\eqref{eq:tauw} it is important to note that only
${\tau_w}_V$ has non-zero mean, whereas integration of equation~\eqref{eq:fik_vel} for $\overline{u}_C$ and $\overline{u}_T$ readily shows
that their integrated contributions vanish as $\boldsymbol{\tau}_C$, $\boldsymbol{\tau}_T$ are both zero at walls.

\begin{table}
\centering
\begin{tabular}{ccccccc}
\hline
\hline
 Case & ${C_f}_V$ & ${C_f}_T$ & ${C_f}_C$ & ${C_f}_V/C_f (\%)$ & ${C_f}_T/C_f (\%)$ & ${C_f}_C/C_f (\%)$ \\
\hline
 A & \num{3.20e-3} & \num{5.36E-3} & \num{6.13E-4} & \num{34.87} & \num{58.44} & \num{6.690} \\
 B & \num{2.03E-3} & \num{5.96E-3} & \num{3.55E-4} & \num{24.30} & \num{71.44} & \num{4.260} \\
 C & \num{8.00E-4} & \num{5.62E-3} & \num{3.40E-4} & \num{11.85} & \num{83.12} & \num{5.030} \\
 D & \num{3.57E-4} & \num{4.79E-3} & \num{3.84E-4} & \num{6.451} & \num{86.60} & \num{6.949} \\
\hline
 A0 & \num{3.20E-3} & \num{6.52E-3} &\num{0} & \num{32.92  }  & \num{67.08} & \num{0} \\
 B0 & \num{2.03E-3} & \num{6.54E-3} &\num{0} & \num{23.65  }  & \num{76.35} & \num{0} \\
 C0 & \num{8.07E-4} & \num{5.84E-3} &\num{0} & \num{12.13  }  & \num{87.87} & \num{0} \\
 D0 & \num{3.57E-4} & \num{4.99E-3} &\num{0} & \num{6.675  }  & \num{93.33} & \num{0} \\
\hline
\end{tabular}
\caption{Contribution of viscous, turbulence and mean convection terms to friction coefficient, as from equation~\eqref{eq:fik}, for DNS A-D and flow cases with suppression of secondary motions A0-D0 (see section~\ref{sec:suppression}).}
\label{tab:fik}
\end{table}
The contributions to the mean friction coefficients are given in table~\ref{tab:fik},
both in absolute terms and as a fraction of the total.
Consistent with physical expectations,
the viscous contribution is observed to decline at increasing Reynolds number with respect to
the turbulent term. The contribution of cross-stream convection is found to be roughly constant
across the explored Reynolds number range, however remaining much less than the turbulence
contribution. Extrapolating the DNS data, we expect the convective contribution to exceed the viscous
one at high enough Reynolds number.

\begin{figure}
 \begin{center}
  \psfrag{x}[t][][1.0]{$z/h$}
  \psfrag{y}[b][][1.0]{$y/h$}
  (a)
  \includegraphics[width=3.8cm,clip]{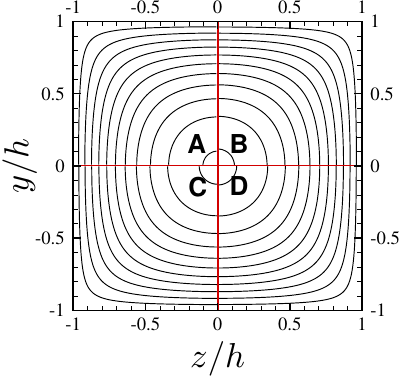}
  (b)
  \includegraphics[width=3.8cm,clip]{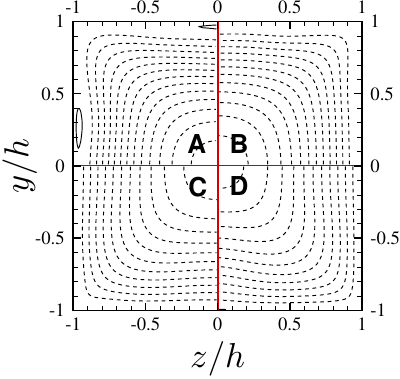}
  (c)
  \includegraphics[width=3.8cm,clip]{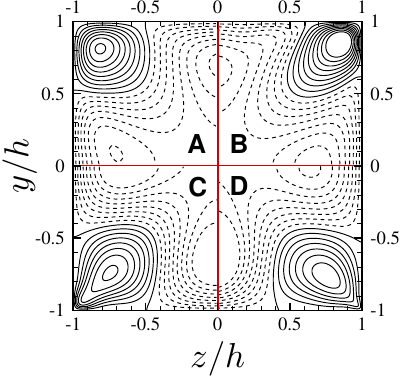}
  \vskip 1em
  \caption{Contributions to mean streamwise velocity $\overline{u}$, 
           each normalized by the corresponding bulk value: viscous $\overline{u}_V/{u_b}_V$ (a), 
           turbulent $\overline{u}_T/\vert {u_b}_T \vert$ (b) and convective $\overline{u}_C/\vert {u_b}_C \vert$ (c),
           as defined in equation~\eqref{eq:fik_vel}.
            Data are shown for flow case A (top left), B (top right), C (bottom left), D (bottom right).
Contour levels are shown for
           $0 \le (.)/ \vert {u_b}_X \vert \le 2 $, in intervals of $2$ in panels (a)-(b) and
           for $-4 \le (.)/ \vert {u_b}_C \vert \le 4$ in intervals of $0.5$ in panel (c) (dashed lines indicate negative values).
          }
  \label{fig:mean_vel_2d}
 \end{center}
\end{figure}
 
Figure~\ref{fig:mean_vel_2d} shows the streamwise velocity fields associated with viscous, turbulent and convective terms. For the sake of clarity, each contribution is normalized with the corresponding bulk value, ${u_b}_X$,
where ${u_b}_T$ and ${u_b}_C$ are both negative, hence providing an additive effect to the 
duct friction (see equation~\eqref{eq:fik}).
The viscous-associated velocity field $u_V$ arises from the solution of a Poisson equation with uniform right-hand-side
(the pressure gradient), hence its shape is identical to the case of laminar flow~\citep{shah_14}, 
only depending on the duct cross-sectional geometry.
The turbulence-associated velocity field $u_T$, shown in figure \ref{fig:mean_vel_2d}(b), 
is everywhere negative and topologically similar to the viscous-associated field, highlighting a retarding effect of turbulence on the bulk flow, with incurred increase of the friction coefficient. 
The velocity field $u_C$ induced by mean convection, shown in figure \ref{fig:mean_vel_2d}(c), 
has a more complex organization. Positive values are found near the duct corners, whereas negative values are found 
near the duct bisectors, the zero crossings being located half-way in between. This finding is consistent with the intuitive expectation
that secondary motions tend to equalize momentum across the duct cross-section, thus quantitatively corroborating
claims made by early investigators~\citep{prandtl_27}. 
Notably, all the distributions shown in figure~\ref{fig:mean_vel_2d} are not qualitatively affected by Reynolds number variation when scaled with respect to their mean integral value, thus showing that changing the 
Reynolds number only changes the relative importance of the three terms, as quantified in table~\ref{tab:fik}.

\begin{figure}
 \begin{center}
  \psfrag{X}[t]{$z/h$}
  \psfrag{Y}[b]{$\tau_w(z)/\tau_w^*$}
  A
  \includegraphics[width=5.5cm,clip]{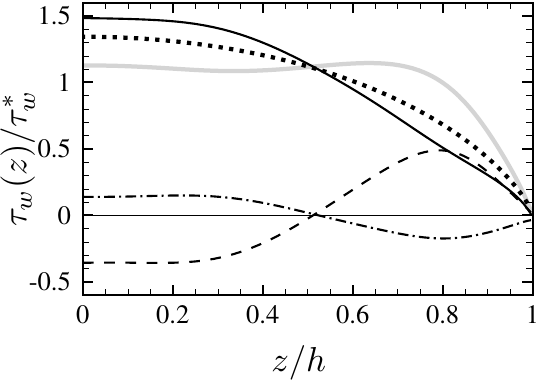} \hskip1.em
  B
  \includegraphics[width=5.5cm,clip]{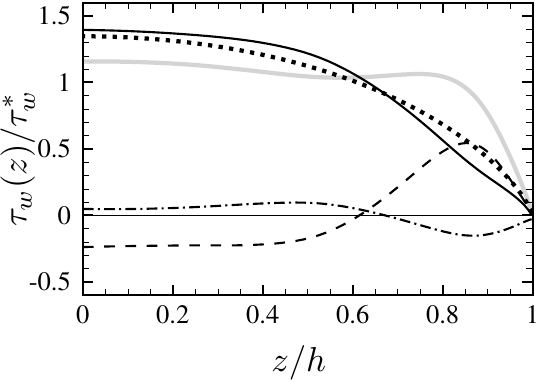}\\
  \vskip 1em
  C
  \includegraphics[width=5.5cm,clip]{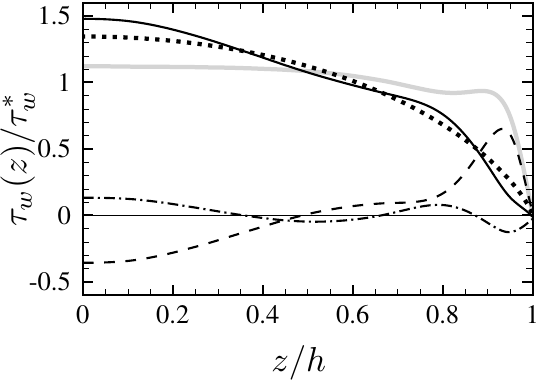} \hskip1.em
  D
  \includegraphics[width=5.5cm,clip]{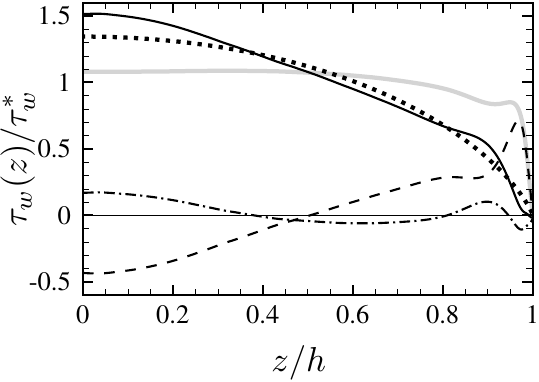}\\
  \vskip 1em
  \caption{Contributions to mean wall shear stress (shown along half duct side): 
           viscous (${\tau_w}_V(z)$, dotted),
           turbulent (${\tau_w}_T(z)$, dash-dotted), 
           viscous + turbulent (${\tau_w}_V(z)+{\tau_w}_T(z)$, solid), 
           mean convection (${\tau_w}_C(z)$, dashed),
           and total ($\tau_w(z)$, thick gray) for flow cases A-D. 
          }
  \label{fig:tauw}
 \end{center}
\end{figure}
The observations made regarding the organization of the streamwise velocity field have direct impact on
the distribution of the local wall shear stress, shown in figure~\ref{fig:tauw}.
The wall shear stress distribution induced by viscous terms (dots) is the same for all cases, and nearly parabolic in shape.
The turbulence terms (dash-dotted lines) have a complex behavior, exhibiting multiple peaks which 
change with the Reynolds number. In general, they yield friction increase toward the duct bisectors, 
and slight attenuation at the corners (compare the solid with the dotted lines), thus 
making the wall friction distributions more nonuniform. Mean convection (dashed lines)
yields large positive peaks in the corner vicinity, whose distance
from the neighboring walls scales in inner units~\citep{pirozzoli_18}, and negative values
around the duct bisector. As a result, the distributions of the wall shear stress (thick gray lines)
tend to be rather flat, especially at high enough Reynolds number. 
This behavior is consistent with recent theoretical analysis based on
inner-outer layer matching arguments~\citep{spalart_18}, which suggest that only uniform wall shear stress 
along the duct perimeter is compatible with the formation of logarithmic layers of the
mean streamwise velocity in the asymptotic high-Re regime.

\begin{figure}
 \begin{center}
  \psfrag{x}[t][]{$(h-r)/h$}
  \psfrag{y}[b][]{$\tau/\tau_w^*$}
%  A
%  \includegraphics[width=3.8cm,angle=270,clip]{../DATA/RETAU150/avrad.eps} \hskip2.em
%  B
%  \includegraphics[width=3.8cm,angle=270,clip]{../DATA/RETAU250/avrad.eps}%\\
%  \vskip 1.5em
  A
  \includegraphics[width=4.0cm,clip]{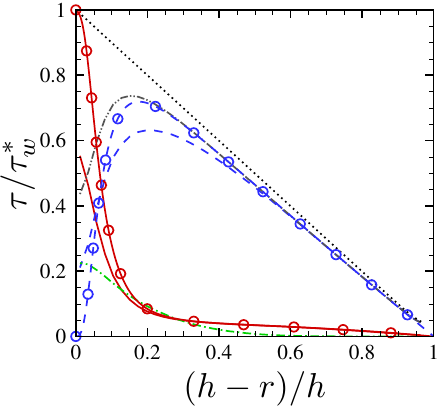} \hskip2.em
  D
  \includegraphics[width=4.0cm,clip]{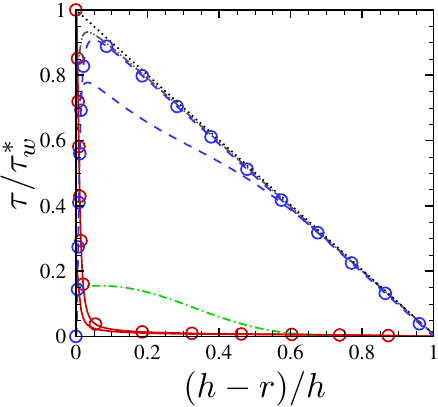}
  \vskip1.em
  \caption{Profiles of azimuthally averaged stresses as a function of the distance from the duct center ($r$), as defined in equation~\eqref{eq:avrad}:
          viscous $\tau_V$, solid; turbulent $\tau_T$, dashed; convective $\tau_C$, dash-dotted; turbulent + convective $\tau_T+\tau_C$, dash-dot-dot; total stress $\tau_t$, dotted. Circles denote pipe flow data~\citep{wu_08} at at $\Rey_{\tau} = 180$ (panel (A)) and $\Rey_{\tau} = 1140$ (panel (D)).}
  \label{fig:avrad}
 \end{center}
\end{figure}
The role of the secondary motions is further analyzed using the
mean momentum balance equation~\eqref{eq:mmb}, averaged over circular shells.
Let $r$, $\theta$ be polar coordinates about the duct center, it is a simple matter to show that 
integrating \eqref{eq:mmb} along the radial direction yields the counterpart of the law of linear
variation of the total stress with the wall distance in canonical flows, namely
\begin{equation}
\tau_t(r) = \tau_V(r) + \tau_C(r) + \tau_T(r) = \tau_w^* r /h, \quad \tau_X = 1/r \int_0^r \rho \left< \nabla \cdot \boldsymbol{\tau}_X \right>_{\theta} \diff \rho, \label{eq:avrad}
\end{equation}
where $r$ is the radial distance from the duct center, 
the local stresses $\boldsymbol{\tau}_X$ are defined in equation~\eqref{eq:mmb}, 
and $<>_{\theta}$ denotes the averaging operator in the azimuthal direction.
Equation~\eqref{eq:avrad} holds for $r \le h$, the upper limit corresponding to the 
condition of tangency of circular shells with the duct walls.
The terms $\tau_X$ represent the viscous, turbulent, and convective stresses 
averaged in the azimuthal direction, and according to equation~\eqref{eq:avrad} their 
sum is supposed to vary linearly with the radial coordinate.
Figure~\ref{fig:avrad} shows the distributions of the averaged stresses as a function of nearest wall
distance, $h-r$.
Mean momentum balance is satisfied with
very good accuracy, as the distributions of the total stress (dotted lines) is very nearly
linear. Interestingly, the distributions of the total stress components are very similar to those
observed in pipe flow, also shown for reference. Specifically, the viscous stress (solid lines)
increases towards the wall, however its value is much less than $\tau_w^*$,
because averaging in the $\theta$ direction also collects
points far from walls where the viscous stress is very small. 
The turbulent stress (dashed lines) has an opposite behavior, being
dominant away from walls, and non-zero for $r \to h$, again on account
of the $\theta$ averaging. The convective stress (dash-dotted lines) is far from negligible, especially
near walls, and its relative importance increases with the Reynolds number. Comparing with
flow statistics from pipe DNS~\citep{wu_08}, we find that when convective and turbulent stresses are taken
together (dash-dot-dotted line), good collapse of the stresses is achieved starting
from a wall distance of about $0.3h$. 
Bearing in mind that nonuniformities in the $\theta$ direction have been removed 
through the averaging procedure, this observation probably points to a universal structure 
of all duct flows, which corroborates the validity of Townsend's
outer layer similarity hypothesis~\citep{townsend_76} also in non-canonical flows.

\section{Suppression of secondary motions} \label{sec:suppression}

 \begin{figure}
  \begin{center}
   \psfrag{X}[t][]{$z/h$}
   \psfrag{Y}[b][]{$y/h$}
   (a)
   \includegraphics[width=5.5cm,clip]{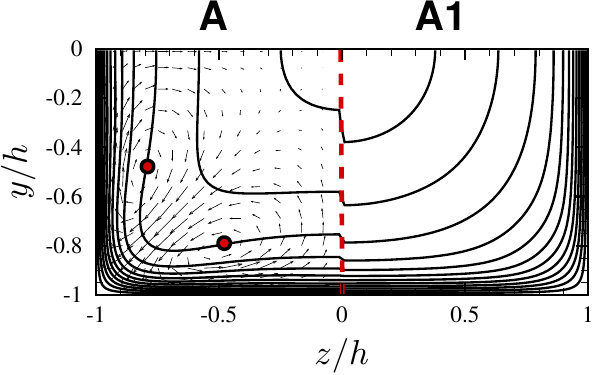}
%  (b)
%  \includegraphics[width=3.5cm,angle=270,clip]{FIGURES/vel_2d_250.eps}\\
%  \vskip 1em
%  (c)
%  \includegraphics[width=3.5cm,angle=270,clip]{FIGURES/vel_2d_500.eps}
   (b)
   \includegraphics[width=5.5cm,clip]{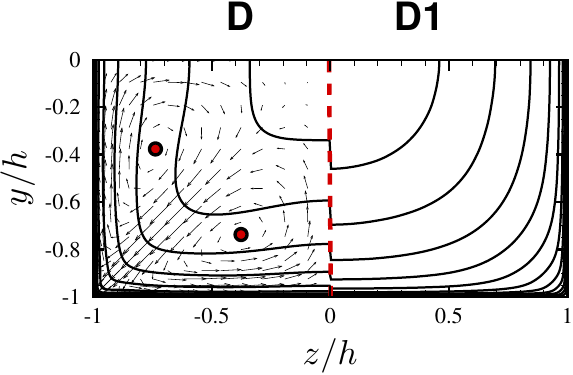}
   \vskip1.em
   \caption{Mean streamwise velocity contours and cross-stream velocity vectors with (left) and without (right)
            secondary motions for flow cases A-A0 (a) and D-D0 (d).
            Thirteen contour levels are shown in the range
            $0 \le \overline{u}/u_b \le 1.3 $. The circles mark the centers of the secondary eddies.
           }
   \label{fig:vel_2d}
  \end{center}
 \end{figure}
Numerical experiments with artificial suppression of the secondary motions (flow cases A0-D0),
are here compared with full DNS (flow cases A-D). 
In figure~\ref{fig:vel_2d} we show mean streamwise
velocity contours in cross-stream planes. Upon suppression of the secondary motions,
the flow does no longer exhibit the typical bulging of the velocity isolines,
because momentum transfer from the duct core towards the corners is inhibited.
In practical terms, this qualitative
change is not accompanied by substantial variation of drag, 
and indeed table~\ref{tab:testcases_duct} shows
modest drag increase for flow cases A0-B0, and
modest drag decrease for flow cases C0-D0. We argue that this non-monotonic behavior is due to
the displacement of the centroids of the secondary eddies in the low-Reynolds-number 
post-transitional regime (see \citet[][figure 18]{pirozzoli_18}).

%\begin{table}
%\centering
%\begin{tabular}{ccccc}
%\hline
%\hline
% Case & ${C_f}_V$ & ${C_f}_T$ & ${C_f}_V/C_f (\%)$ & ${C_f}_T/C_f (\%)$ \\
%\hline
% A0 & $3.20E-3$ & $6.52E-3$ & $32.92  $  & $67.08$ \\
% B0 & $2.03E-3$ & $6.54E-3$ & $23.65  $  & $76.35$ \\
% C0 & $8.07E-4$ & $5.84E-3$ & $12.13  $  & $87.87$ \\
% D0 & $3.57E-4$ & $4.99E-3$ & $6.675  $  & $93.33$ \\
%\hline
%\end{tabular}
%\caption{Contribution of viscous and turbulence terms to friction coefficient, as from equation~\eqref{eq:fik}, for flow cases with suppression of secondary motions.}
%\label{tab:fik_novw}
%\end{table}

The analysis of the contributions to the total friction carried out in Section~\ref{sec:role} 
has been repeated for the flow cases without secondary motions, and the results are reported in table~\ref{tab:fik} 
for flow cases A0-D0.
It appears that the viscous contribution is very nearly the same as in the baseline
cases, and the absence of convection is almost entirely compensated by increase
of the turbulent contribution, in such a way that the overall variation of friction is quite small.

\begin{figure}
 \begin{center}
  \psfrag{X}[t]{$z/h$}
  \psfrag{Y}[b]{$\tau_w(z)/\tau_w^*$}
  A0
  \includegraphics[width=5.5cm,clip]{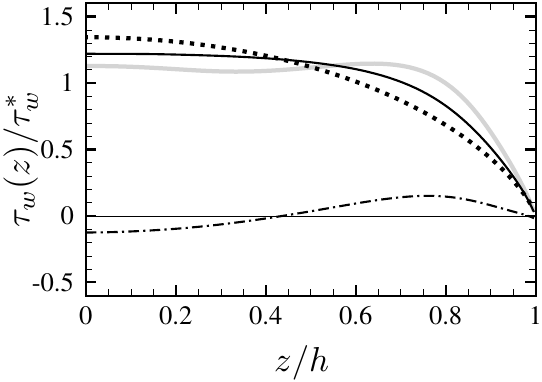} \hskip1.em
  B0
  \includegraphics[width=5.5cm,clip]{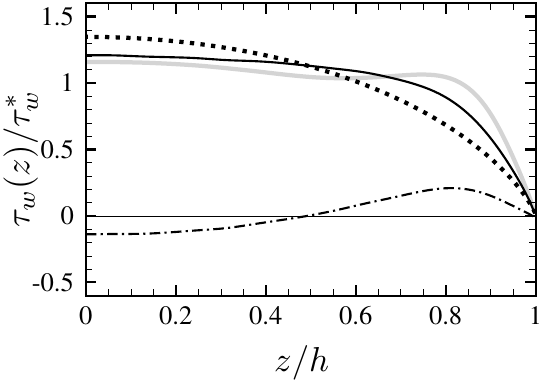}\\
  \vskip 1em
  C0
  \includegraphics[width=5.5cm,clip]{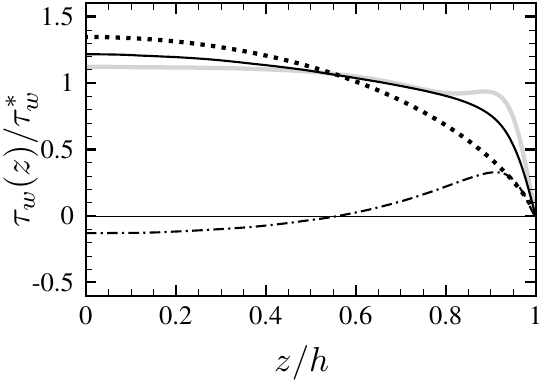} \hskip1.em
  D0
  \includegraphics[width=5.5cm,clip]{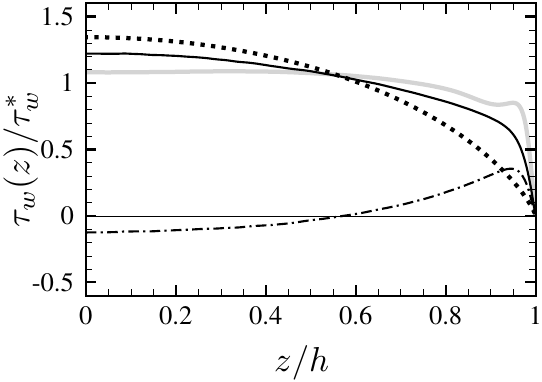}\\
  \vskip 1em
  \caption{Contributions to mean wall shear stress (shown along half duct side): 
           viscous (${\tau_w}_V(z)$, dotted),
           turbulent (${\tau_w}_T(z)$, dash-dotted), 
           and total ($\tau_w(z)$, solid) for flow cases A0-D0. 
           The wall shear stress for flow cases A-D is
           also shown for comparison with thick gray lines.
          }
  \label{fig:tauw0}
 \end{center}
\end{figure}

The distributions of the wall shear stress for flow cases A0-D0 are shown in figure~\ref{fig:tauw0}. 
Overall, the wall shear stress (solid lines) is less flat than in flow cases A-D (gray lines), and it does no longer
exhibit the typical near-wall peaks caused by mean convection. 
Comparing with figure~\ref{fig:tauw} shows that the viscous contribution obviously retains the same shape, 
whereas the turbulence contribution is qualitatively modified by suppression of the secondary motions. 
In particular, ${\tau_w}_T$ (dash-dotted lines) has reversed
sign, becoming positive towards the corners and negative at the duct mid-side. 
This observation is consistent with the fact that the associated velocity field (not shown) 
has a structure similar to the convection term in figure~\ref{fig:mean_vel_2d}(c), thus 
bringing high-momentum fluid towards the corners.

\begin{figure}
 \begin{center}
  \psfrag{X}[t][][1.2]{$y^+$}
  \psfrag{Y}[b][][1.2]{$u^+$}
  \psfrag{x}[t][][0.5]{$z/h$}
  \psfrag{y}[b][][0.5]{$y/h$}
%  (A)
%  \includegraphics[width=4.5cm,angle=270,clip]{FIGURES/vel_all_150.eps}
%  (A0)
%  \includegraphics[width=4.5cm,angle=270,clip]{FIGURES/vel_all_150_novw.eps}\\
%  \vskip 1em
%  (B)
%  \includegraphics[width=4.5cm,angle=270,clip]{FIGURES/vel_all_250.eps}
%  (B0)
%  \includegraphics[width=4.5cm,angle=270,clip]{FIGURES/vel_all_250_novw.eps}\\
%  \vskip 1em
%  (C)
%  \includegraphics[width=4.5cm,angle=270,clip]{FIGURES/vel_all_500.eps}
%  (C0)
%  \includegraphics[width=4.5cm,angle=270,clip]{FIGURES/vel_all_500_novw.eps}\\
%  \vskip 1em
  D
  \includegraphics[width=5.5cm,clip]{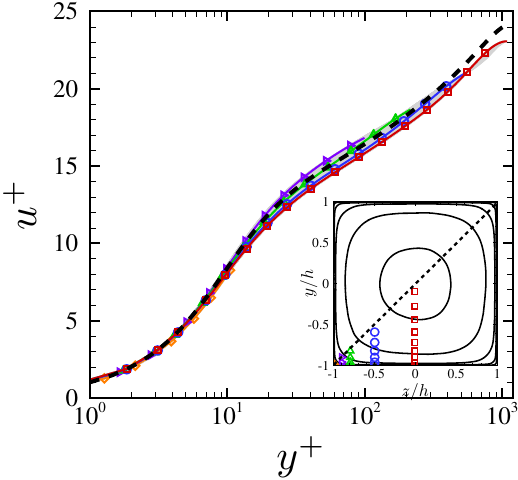} \hskip1.em
  D0
  \includegraphics[width=5.5cm,clip]{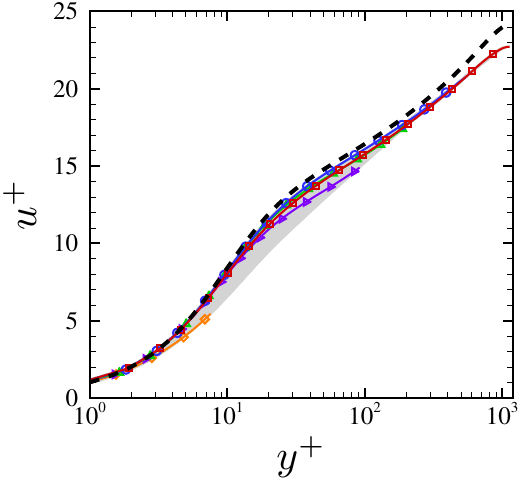}\\
  \vskip 1em
  \caption{Mean streamwise velocity profiles along the wall-normal $y$ direction,  given in local wall-units at all $z$ up to the corner bisector, for flow case D and D0. Representative stations along the bottom wall
are highlighted, namely $(z+h) = 15 \delta_v^*$ (diamonds), $(z+h)/h = 0.1$ (right triangles), $(z+h)/h = 0.25$
(triangles), $(z+h)/h = 0.5$ (circles), $(z+h)/h = 1$ (squares), see inset in panel (a). The dashed lines denote profiles from DNS of pipe flow flow at $Re_\tau=1140$~\citep{wu_08}.
          }
  \label{fig:mean_vel}
 \end{center}
\end{figure}

The inner-scaled wall-normal mean velocity profiles are shown in
figure~\ref{fig:mean_vel} for the highest Reynolds number cases. For the sake of clarity, the 
profiles are scaled in local wall units at each $z$ location along the bottom wall, and shown up to the corner bisector $y = z$ (see
inset in the left panel), where velocity has local maxima. The mean velocity profiles 
for circular pipe flow~\citep{wu_08} are also included for comparison. 
As pointed out by \citet{pirozzoli_18}, local wall scaling yields excellent collapse of
the velocity distributions up to the corner bisector, thus supporting the 
robustness of the law-of-the-wall in wall-bounded turbulence. 
However, removing the secondary motions
(right panel) yields greater data scatter in the buffer layer, 
and deviations from the pipe velocity profile are consequently larger also in the outer layer. 
In this respect we again recall that the analysis of~\citet{spalart_18}
shows that a log law can only be achieved in ducts with complex cross section if the wall shear
is uniform. 
Hence, the observed greater universality of the inner-scaled velocity profiles in the
presence of secondary motions is likely to be the side result of greater uniformity in the wall shear
stress distributions.
In this sense, we may further state that secondary motions contribute to establishing universality of the flow statistics.

\section{Conclusions}

%We have carried out DNS of square duct flow up to friction Reynolds number $\Rey_\tau \approx 1000$, to
%shed light on the effect of Reynolds number on secondary motions, and clarify their role in the
%mean flow organization. For that purpose, we have derived a generalized version of the classical 
%FIK identity, which we have used to isolate and quantify the effect of viscous, 
%convective and turbulent terms on the mean
%velocity field, on the wall shear stress distributions, and on the average friction coefficient. Numerical
%experiments have also been carried out whereby secondary motions are artificially suppressed
%through suitable forcing of the Navier-Stokes equations.
Based on the results of this study, we are in a position to judge about the actual effect of secondary motions on the
flow statistics, thus providing quantitative answers to a frequently debated subject.
As regards the bulk flow properties, we find that 
secondary motions contribute by as much as $6\%$ to the
mean duct friction coefficient, hence their suppression may in principle yield some relief in terms
of reduced pressure drop. This expectation is not realized in practice, and we find that artificially suppressing the
secondary motions changes the structure of the turbulence terms in a partially compensating fashion.
Hence, drag increase is found at low Reynolds number, and drag reduction of no more than $3\%$, 
at sufficiently high Reynolds number. 
Notably, this drag reduction is comparable to that obtained with a change of duct shape from square to circular.
Indeed, assuming that friction is controlled by the Reynolds number based on the hydraulic diameter, and considering for simplicity
Blasius friction law, namely $C_f = 0.079 \Rey_D^{-1/4}$, it is easily found that the ratio of the friction
coefficient in a square duct as compared with a circular pipe with the same area is $(2/\sqrt{\pi})^{1/4} \approx 1.03$.

As regards the general flow organization, we are able to quantitatively confirm that the main
role of the secondary motions is to bring high-momentum fluid from the side bisectors towards
the corners, hence tending to compensate the momentum defect at the duct corners 
and to make for a fuller mean velocity profile, thus counteracting nonuniformities induced by the viscous and
turbulence terms. As a result, the distributions of the wall shear stress
also tends to be more uniform than would be obtained if only the viscous
and turbulence terms are retained. 
Artificial suppression of the secondary motions changes the qualitative structure of the
turbulence terms, which then assume a uniforming role similar to mean convection. 
However, the wall shear stress distributions are less flat than in the
case of the full DNS. As a result, DNS including the secondary motions 
tend to have greater universality of the wall-normal mean velocity profiles,
which follow with good fidelity the canonical log law.

A further outcome of the present analysis is that the azimuthally-averaged mean momentum budget terms
reveals a structure similar to that found in circular pipe flow, thus supporting extended validity
of Townsend's similarity hypothesis also for flows in ducts with complex shape.
In the presence of secondary motions, mean convection provides a contribution comparable to the turbulent
stresses, especially near the walls, which however adds to the turbulent contribution yielding
an effective averaged stress distribution which is very similar to pipe flow.

%In summary we find that, although secondary motions in square ducts are weak ($1-2\%$ of the bulk velocity),
%they have a significant impact on the flow organization, especially as they tend to yield flatter
%distributions of the wall shear stress, thus also making the wall-normal velocity distributions
%universal and close to the log law. In global terms, their effect is rather small, and we estimate that
%secondary motions contribute to mean friction by no more than a few percent.

\begin{acknowledgements}
We acknowledge support from J. Jimenez for hosting us during the Third Madrid Turbulence Workshop, 
funded by the COTURB ERC project.
We further acknowledge that the data reported in this paper have been computed using the PRACE Research Infrastructure resource MARCONI based at CINECA, Casalecchio di Reno, Italy.
\end{acknowledgements}

%\section*{References}
%\clearpage
%\bibliographystyle{iopart-num}
%\newcommand{\newblock}{}
\bibliographystyle{jfm}

\bibliography{references}

\end{document}